\documentclass[12pt]{article}
\usepackage{amsmath}
\usepackage{colortbl}
\usepackage{xcolor}
\usepackage{graphicx}
\usepackage{caption}
\usepackage{lipsum} % To generate dummy text
\usepackage{geometry}
\usepackage{tabularx}

\title{\textbf{A Hybrid Quantum-Chaos Theory Approach to Image Encryption Using Reservoir Computing}}
\author{Naheen Mohd. Kadir \\ naheen.kadir@gmail.com
	\\ Department of Computer Science \& Engineering
	\\ RUET, Rajshahi, Bangladesh}
\date{June, 2025 (Undergrad Thesis)}
\begin{document}
	
	\maketitle
	\noindent\rule{\textwidth}{0.4pt}  % This creates a horizontal line
	\vspace{0pt}
	\begin{abstract}
		This research presents a novel hybrid image encryption system that combines quantum cryptography, chaos theory and reservoir computing to address the limitations of conventional encryption methods. With the rapid advancements in quantum computing, traditional systems like RSA, DHKE (related to prime number factorization) are vulnerable to quantum attacks i.e. Shor’s algorithm, Grover’s algorithm. In response, this study proposes a hybrid solution that uses quantum cryptographic protocols, particularly the E91 protocol, to generate secure, eavesdrop-proof keys through quantum entanglement. The integration of chaos theory, specifically the Lorenz hyper-chaotic system, enhances the encryption system by adding unpredictability and sensitivity to initial conditions, making it more resistant to both classical and quantum-based attacks. Reservoir computing is used to improve computational efficiency, enabling faster and more effective encryption and decryption processes. The system achieved encryption times as low as 0.0296s and decryption times as low as 0.0164s for 128×128 images, with MAE reduced for 300–600 node networks, and NPCR and UACI above 99.6\% and 0.49, respectively, across various image sizes and configurations. By combining quantum cryptography, chaos theory and reservoir computing, this approach offers both enhanced security and practical feasibility for image encryption in the age of quantum computing.
		\newline
		
	\end{abstract}
	\section{Introduction}
	The rapid development of quantum computing \cite{booknielsen2010quantum} and chaos theory \cite{chaostheoryandapparticle} has introduced paradigm-shifting in cryptography and data security. Encrypting images is a very important way to keep private visual data safe from people who want to access them illegally over a network. Traditional encryption methods, on the other hand, have a lot of issues \cite{appliedcarticle}, mostly because they rely on formulas that are easy to figure out \cite{qvur10439158} by quantum computer. Classical types of encryption are simple to get around. We can see two examples, RSA \cite{rsa10.1145/359340.359342} and DHKE \cite{dhke1055638}. They are safe because it is hard to factor big prime numbers. If we have a powerful enough quantum computer, Shor's algorithm \cite{doi:10.1137/S0097539795293172} can de-touch prime numbers from multiplication of big prime numbers in a very short amount of time. In other words, it can get around RSA cryptography. In the same way, Grover's algorithm \cite{grover10.1145/237814.237866} can weaken traditional encryption methods, making it easier for brute-force attacks to find valid keys.
	Keys that are safe and can't be read by anyone else are made with quantum encryption protocols \cite{griffiths_introduction_2018}, such as the E91 protocol \cite{e91PhysRevLett.67.661}. On the other hand, chaos theory has done a great job of making cryptography safer \cite{chaosrev}. This is because it is usually unpredictable and depends a lot on how it initiates. This convergence of quantum and chaotic systems may offer a feasible pathway toward quantum-resistant image encryption methodologies.
	
	The main objective of this research is to find ways to fix the problems with current image encryption methods. As quantum computing gets better, encryption methods that rely on breaking down prime numbers are becoming less secure. Chaotic dynamical systems, especially the Lorenz attractor model \cite{lorenz1963deterministic}, are a very interesting option because they can create complex, non-linear patterns that are very sensitive to initial conditions. Because they are so unpredictable, these systems are very hard to break with regular cryptanalysis methods. They are also more secure against both classical and quantum attacks.
	Reservoir Computing \cite{rcenc1JIANG2024124913}\cite{esnjaeger2001echo} also makes it easier to work with complex, high-dimensional data by speeding up and improving the processes of encryption and decryption. The goal of this research is to come up with a hybrid method that uses these modern computing techniques to make image encryption safer.
	The E91 protocol is very safe to generate keys. Chaos-based systems can then make the process even more unpredictable with addition of E91 protocol. Putting all of these components together will allow the technique for encrypting images work much better. It will also be able to deal with both classical and quantum threats that might come up in the future.
	
	The purpose of this research is to make a hybrid encryption system that uses the E91 protocol for secure quantum key distribution, the Lorenz system for producing random chaotic sequences and and reservoir computing for quick and safe processing of image data. The process makes sure that the encryption is powerful enough to keep anybody from break in, which makes it safer and more effective.
	
	\section{Existing Research and Developments}
	
	Quantum cryptography, chaos-based encryption and reservoir computing all coming together could change the way image encryption works. Quantum key distribution (QKD), chaos theory, and reservoir computing are all fascinating technologies on their own, but merging them into a single, multi-layered encryption system is an attractive research possibility. 
	
	Quantum cryptography, especially Quantum Key Distribution (QKD) \cite{quantcrypPanda2025} and Quantum Machine Learning (QML) \cite{qmlKedia2025}, has a quantum advantage in keeping information safe through quantum entanglement \cite{booknielsen2010quantum}, like the E91 QKD protocol \cite{e91PhysRevLett.67.661}. QKD provides unbreakable security in perfect conditions, but it is still hard to combine with other encryption schemes since it is hard to combine quantum and classical channels. Recent research has focused on using QKD with classical algorithms like AES \cite{aesbook} or DES \cite{desschneier2007applied} to encrypt steganographic image \cite{SYKOT2025167}. But the usage of classical algorithm which follows the mathematical rules is vulnerable to quantum computer.
	
	Chaos-based encryption systems, utilizing chaotic phenomena such as the Lorenz attractor \cite{lorenz1963deterministic}, have been explored for their potential in secure image encryption \cite{chaosrevarticle}. The inherent unpredictability and sensitivity to initial conditions make them resilient to traditional cryptanalysis. However, chaos-based encryption systems are susceptible to new forms of attack, such as those leveraging Machine Learning techniques \cite{mlatqchaos10551181}. Furthermore, chaotic map systems like the Arnold map and Chen's chaotic system are computationally expensive for large images \cite{guanarticle}. Nevertheless, deep learning-based chaotic systems \cite{dlchaosZHOU2025111017} show promise in improving the efficiency and robustness of image encryption.
	
	Reservoir computing, particularly echo state networks (ESNs) \cite{esnjaeger2001echo}, offers an efficient solution for speeding up encryption and decryption processes, particularly in image encryption \cite{rcenc1JIANG2024124913}. Reservoir computing models are faster than traditional neural networks, but challenges remain in scaling them to handle larger images while maintaining security. Further research is needed to optimize reservoir computing architectures for high-dimensional image data and ensure that security is not compromised.
	
	Integrating these technologies—quantum cryptography, chaos-based encryption, and reservoir computing—into a hybrid system could provide several key benefits. By combining quantum key distribution with chaos-based systems and the computational power of reservoir computing, we can create a system that not only resists quantum-based attacks but also enhances computational efficiency and security. For instance, the unpredictability of chaos-based systems could be used to generate random keys, while quantum cryptography ensures the secure exchange of those keys. Reservoir computing might make the encryption process faster, which would make it useful for applications that need to work in real time.
	
	The main research opportunity is to create an integrated encryption system that uses QKD (with the E91 protocol), chaos-based encryption (like the Lorenz system), and reservoir computing all at the same time. This mixed method could help solve a number of important problems. The proposed system leverages the E91 protocol to keep assaults like Shor's and Grover's algorithms, which are hard for quantum computers to break. It leverages reservoir computing to speed up and improve the efficiency of encryption and decryption compared to most chaos-based solutions. Using both chaotic sequence generation and quantum key distribution makes guarantee that attacks on both classical and quantum cryptography can't get through. This means that the system will be safe for a long time.
	
	Filling up these research gaps and combining these methods could lead to the creation of a very secure and effective image encryption system that can protect against both traditional and quantum attacks while taking advantage of the processing power of modern technology.
	
	\section{Preliminaries}
	\subsection{E91 Protocol}
	Artur Ekert made the E91 protocol in 1991. The E91 protocol uses quantum entanglement and a Bell-state \cite{bellGISIN19981} test to find any eavesdroppers. The security of the protocol depends on the fact that quantum entanglement is not local. It makes sure that any attempt to intercept or measure the quantum data would cause problems for the system \cite{booknielsen2010quantum} and let the people who are communicating know.
	The E91 Protocol works as follows:
	\begin{itemize}
		\item \textbf{Entangled Particle Pairs:} 
		At the core of the E91 protocol is the use of entangled particle pairs. A single source emits pairs of entangled photons, which are typically polarized. The entangled particles are created in a Bell state \cite{bellGISIN19981}, particularly in the singlet state. The singlet state is described as follows:
		\[
		| \Psi \rangle = \frac{1}{\sqrt{2}} (|01 \rangle - |10 \rangle)
		\]
		This state implies that if one particle is measured to be in a particular polarization state (say, vertical or horizontal), the other particle will be measured in the corresponding state (opposite polarization). The crucial feature of entanglement is that the outcome of a measurement on one particle instantaneously determines the state of the other, even if the particles are far apart.
		
		After the particles are entangled, one photon is sent to Alice, and the other is sent to Bob. This separation allows the two parties to perform measurements on their respective photons independently.
		
		\item \textbf{Measurement Process:}
		Alice and Bob each measure the polarization of their received photons. To ensure privacy and security, they do not use fixed bases for their measurements. Instead, they randomly choose from a set of measurement bases. The choice of measurement bases is according to the CHSH test.
		\item \textbf{CHSH Test:} The Clauser-Horne-Shimony-Holt inequality is an important part of the E91 strategy for finding eavesdroppers. The CHSH test looks to see if Alice and Bob's values remain connected even though they are based on different units.
		
		The CHSH inequality is defined as:
		\[
		S = E(A, B) + E(A, B') + E(A', B) - E(A', B')
		\]
		Where $E(A,B)$ represents the correlation between Alice’s and Bob’s measurements using bases $A$ and $B$, respectively. The measurements fail to follow the CHSH inequality if the number of $S$ is higher than a certain level.
		\item \textbf{Measurement Bases:}
		The below shows the measurement bases of Alice and Bob:
		\begin{itemize}
			\item Alice’s measurement bases:
			\begin{itemize}
				\item $B_1$ (horizontal/vertical polarization)
				\item $B_2$ (diagonal/anti-diagonal polarization)
				\item $B_3$ (circular polarization)
			\end{itemize}
			\item Bob’s measurement bases:
			\begin{itemize}
				\item $B_1$ (horizontal/vertical polarization)
				\item $B_2$ (diagonal/anti-diagonal polarization)
				\item $B_3$ (circular polarization)
			\end{itemize}
		\end{itemize}
		
		\item \textbf{Basis Discussion and Eavesdropper Detection:} 
		After measurement process, Alice and Bob publicly discuss their bases but they do not share the actual results of their measurements. This public discussion is essential because it allows them to check whether they used the same bases during measurement. If they are measured using the same bases, their results will be correlated, and they can use these correlated results to generate a shared secret key.
		However, if an eavesdropper (Eve) tries to intercept the communication, her intervention will disturb the quantum states, causing discrepancies in the correlation between Alice and Bob’s results. These discrepancies can be detected, and Alice and Bob can discard the compromised bits. This detection capability forms the basis for the security of the E91 protocol.
	\end{itemize}
	\subsection{Chaos Based Encryption}
	Chaotic systems have long been explored for cryptographic applications due to their sensitivity to initial conditions and their ability to produce seemingly random behaviors, ideal for encryption. Two notable chaos-based encryption techniques are chaotic map-based encryption and deep learning-enhanced chaotic encryption.
	
	\textbf{Chaotic Map-Based Encryption:} Guan et al. \cite{guanarticle} proposed a two-phase encryption method using spatial permutation and value spread. The first phase involves \textbf{pixel shuffling} using the Arnold Cat Map, which disrupts pixel relationships by applying a linear transformation. This transformation is controlled by secret parameters $p$, $q$, and the number of iterations $M$. In the second phase, the \textbf{pixel values} are encrypted using Chen’s chaotic system \cite{chenarticle}, where the chaotic outputs are XOR-ed with pixel values to produce the cipher image. This method offers a large key space ($10^{42}$ for 64-bit precision), high sensitivity to initial conditions, and a uniform histogram distribution that defends against statistical attacks. However, it has limitations, including the periodicity of the Arnold Cat Map, which could potentially be exploited by attackers, and the normalization of chaotic outputs to 8-bit pixel values, which reduces randomness and weakens security. Additionally, the method is vulnerable to chosen-plaintext attacks.
	
	\textbf{Deep Learning-Enhanced Chaotic Encryption:} To address the weaknesses of traditional chaos-based methods, Panwar et al. \cite{panwarsystems11010036} introduced \textbf{EncipherGAN}, a deep learning-based encryption system that integrates chaotic principles with neural networks. This system uses a cycle-consistent generative adversarial network (cycle-GAN) for both encryption and decryption. The encoder-decoder architecture, enhanced with residual blocks, ensures stable feature transformation and preserves image structure. A convolutional discriminator network evaluates cipher images, ensuring they resemble real encrypted data. The encryption quality is further refined using a loss function that combines adversarial loss, cycle-consistency loss, and Structural Similarity Index Metric (SSIM) optimization. The approach offers advantages such as a larger key space, high-quality decryption (PSNR ~40 dB), and resistance to statistical and plaintext attacks. However, the method faces challenges such as significant computational overhead during network training and the need for diverse datasets to ensure robustness.
	
	Both methods leverage chaotic principles for strong encryption but have different strengths and limitations, with deep learning-enhanced methods offering more advanced security features at the cost of higher computational complexity.
	\subsection{Lorenz Hyper-chaotic System}
	The classical form of the Lorenz system \cite{lorenz1963deterministic} consists of three differential equations.It show how the system's state changes over time. But here, we will use a hyperchaotic version of the Lorenz system \cite{hyperchaosWANG20083751}. It adds a fourth state variable \(w(t)\), which makes the system more hyper-chaotic, better than only chaotic and better for cryptography obviously.The four equations governing the system are:
	
	\[
	\frac{dx}{dt} = a(y - x) + w
	\]
	\[
	\frac{dy}{dt} = c x - y - x z
	\]
	\[
	\frac{dz}{dt} = x y - b z
	\]
	\[
	\frac{dw}{dt} = - y z + r w
	\]
	
	Where:
	\begin{itemize}
		\item \(x(t), y(t), z(t), w(t)\) represent the four variables of the system (state variables).
		\item \(a, b, c, r\) are parameters that define the behavior of the system.
	\end{itemize}
	\subsection{Reservoir Computing}
	Reservoir Computing (RC) \cite{rcenc1JIANG2024124913} \cite{rc2Paul2025} is now a powerful way to model systems that change over time and make predictions based on time series \cite{time1WYFFELS20101958} \cite{time29127499}. It is a well-known and effective recurrent neural network (RNN) design that can do things like predict chaotic time series with very little training data \cite{esnjaeger2001echo}. RC works by adding more dimensions to the input data in a nonlinear way using a set of randomly initialized reservoirs. This is done before training just a linear readout layer. This design saves a lot of money on computing power compared to regular RNNs, which require training the whole network.
	Reservoir computing is a popular choice for people who want cheaper or less data-centric model  than regular deep learning models. It also works well for quick cryptography tasks because it only need to train the output ridge, can model complicated patterns, and is not easy to break with normal cryptanalysis. This section will first explain what RC is and how it works. Next, it will talk about how it can be used for safety.
	
	There are three main parts that make up RC that work together to process data quickly and correctly:
	
	1. \textbf{Input Layer ($\mathbf{W}_{in}$)}:
	\begin{itemize}
		\item The input layer is a randomly initialized matrix. It maps the input data $\mathbf{u}(t) \in \mathbf{R}^M$ to a higher-dimensional space and that is called the reservoir. This layer modifies the input first, which makes the image more complicated. The entries of $\mathbf{W}_{in}$ are typically drawn from a uniform distribution, typically in the range $[-k, k]$, where $k$ is a tunable hyperparameter that controls the scale of the transformation.
	\end{itemize}
	
	2. \textbf{Reservoir Dynamics}:
	\begin{itemize}
		\item The reservoir’s state $\mathbf{r}(t + \Delta t)$ is updated via a nonlinear transformation. The state evolves according to the following equation:
		$$
		\mathbf{r}(t + \Delta t) = \tanh(\mathbf{W}_{in}\mathbf{u}(t) + \mathbf{A}\mathbf{r}(t)),
		$$
		where $\mathbf{A}$ represents the adjacency matrix of the reservoir, and the $\tanh$ function acts as the activation function. The reservoir network is fixed and random, and its primary function is to nonlinearly expand the input data into a high-dimensional state space that captures the complex dynamics of the input.
	\end{itemize}
	
	3. \textbf{Readout Layer ($\mathbf{W}_{out}$)}:
	\begin{itemize}
		\item The readout layer is a linear layer that is trained using a simple regression technique, typically ridge regression, to map the reservoir states to the desired output. The optimization minimizes the error between the predicted output and the actual target. The optimization objective is given by:
		$$
		\mathbf{W}_{out} = \arg\min \left( \sum_{t=1}^{\tau} ||\mathbf{W}_{out}\mathbf{r}(t) - \mathbf{u}(t)||^2 + \beta ||\mathbf{W}_{out}||^2 \right),
		$$
		where $\beta$ is a regularization parameter that prevents overfitting and ensures the stability of the learned weights.
	\end{itemize}
	
	This approach, where only the readout layer is trained, significantly reduces computational complexity compared to traditional RNNs, which require the training of all network weights.
	\subsection{Reservoir Computing in Encryption}
	Reservoir Computing is suitable for cryptographic applications due to its fast computing mechanism. In RC-based encryption, the high-dimensional reservoir unit makes it hard to figure out how the data was protected by what. Two different kinds of RC-based cryptography were studied:
	
	\textbf{Prediction Mode:}
	Proposed by Ramamurthy et al. \cite{ramaDBLP:journals/corr/RamamurthyBBW17}, this mode use a shared secret key $\mathbf{k}$ with identical RC networks between the communicating parties. In this mode:
	\begin{itemize}
		\item Alice encrypts an image by training the reservoir states by predicting the image pixels. This process uses the chaotic behavior of the reservoir by performing pixel-wise predictions. Here, the key $\mathbf{k}$ guides the prediction process of reservoir.
		\item Bob performs decryption by applying the learned $\mathbf{W}_{out}$ reservoir states.Then is is reconstructed using the same key $\mathbf{k}$. By iteratively applying the readout layer, the method can recover the original image.
	\end{itemize}
	
	\textbf{Separation Mode:}
	Introduced by Krishnagopal et al. \cite{krisharticle}, this mode treats the secret key as input $\mathbf{X}$ and the image as the target $\mathbf{Y}$. In this approach:
	\begin{itemize}
		\item The sender trains the readout layer $\mathbf{W}_{out}$ of the reservoir to separate the target image $\mathbf{Y}$ from the key $\mathbf{X}$. The method utilizes reservoir computing to separate the image data from the secret key using the complex dynamics of reservoir.
		\item The receiver decrypts the image through a single matrix multiplication: $\mathbf{Y} = \mathbf{W}_{out}\mathbf{R}$, where $\mathbf{R}$ is the reservoir state at the receiver’s end.
	\end{itemize}
	
	This method has been shown to make encryption and decoding much faster, with speeds of up to 7.6 MB/s. This works a lot faster than prediction-based RC methods and old-fashioned chaos-based encryption methods \cite{kol10191083}.
	\section{Proposed Architecture}
	\subsection{Overview of Full Proposed Hybrid Approach}
\begin{figure}[h!]
		\centering
		\includegraphics[width=0.9\textwidth]{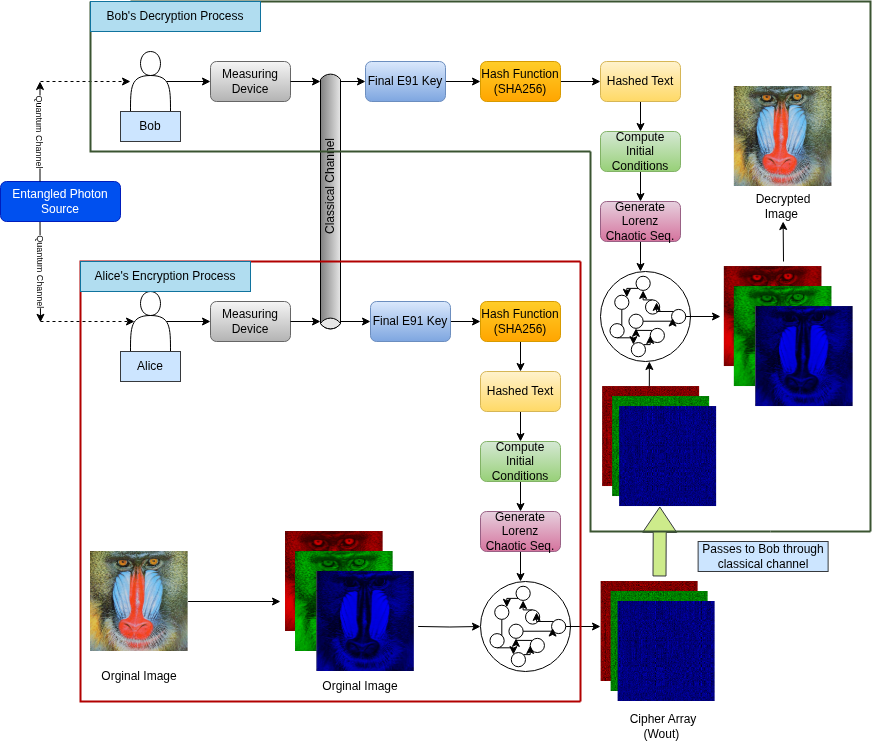} 
		% Replace with the path to your model diagram image
		\caption{Overview of the Proposed Hybrid Image Encryption Model}
		\label{fig:hybrid_model}
	\end{figure}

	The proposed hybrid encryption system uses reservoir computing to integrate quantum key distribution with chaotic systems. It encrypts and decrypts images quickly and safely. This system employs quantum cryptography to safely produce and send keys, hashing and chaos theory to make things very unpredictable because of its randomness, and reservoir computing to ensure that encryption and decryption take place quickly by just using the output layer. Fig \ref{fig:hybrid_model} shows a summary of how the method works.
	
	\subsubsection{ \textbf{Alice's Encryption Process}}
	The suggested encryption method starts with Alice and Bob sharing entangled photons using the E91 Quantum Key Distribution (QKD) protocol. This makes sure that the keys are sent securely through quantum entanglement. Alice then hashes the key using SHA-256 to make a unique cryptographic key. This key is used to compute the initial conditions. These starting conditions are used to set up a Lorenz-based hyper-chaotic system that makes pseudo-random sequences that are very sensitive to starting conditions, which makes them hard to predict.
	
	When we encrypt an image, we break it up into RGB channels and then use the chaotic sequence and the encryption key to treat each channel separately. After that, Bob gets the encrypted image (Wout) through a regular channel. The encrypted image doesn't include any original pixel data; instead, it is shown as output weight vector in reservoir computing, which makes the network perform better and keeps it safer.
	\subsubsection{ \textbf{Bob's Decryption Process}}
	Bob utilizes the same E91 process to make the final decryption key by measuring the entangled photon pair with quantum technology. The key that Bob made is the same as the one that Alice used. Next, the key goes through SHA-256 hashing, which makes sure that it is correct and creates a hashed text that will be used to decrypt it.
	Bob then takes the hashed key to figure out the starting conditions and uses those to make the Lorenz hyper-chaotic sequence that goes with them, just like Alice did. Bob uses the created chaotic sequence and the decryption key to reverse the operations on the encrypted image (Wout) and get the original RGB channels back. To put the original image back together, these channels are processed in the opposite sequence.
	Finally, after the decryption process is finished, the RGB channels are combined, and Bob gets the decrypted image, which is precisely the same as the original image.
	\subsection{E91 Protocol for Secure Key Distribution}
	The E91 protocol uses quantum entanglement to make sure that Alice and Bob can safely share keys without anybody else being able to break in. Entangled photon pairs are made, and each person measures the polarization of their photon in a randomly determined basis, such as diagonal or rectilinear. The correlation of measurement results creates a shared secret key, and Bell's theorem violations will find any eavesdropper, ensuring security. The key generation process is formalized as follows:
	\begin{equation}
		S = \langle A_1 B_1 \rangle + \langle A_1 B_2 \rangle + \langle A_2 B_1 \rangle - \langle A_2 B_2 \rangle,
	\end{equation}
	where \( S \) is the Bell parameter, and \( A_i, B_i \) are Alice's and Bob's measurement outcomes. If \( |S| > 2 \), the system is secure.
	\subsection{Calculating Initial Conditions}
	After successfully shared the key to Alice and Bob using E91 protocol, we hased the key using SHA256. After getting the hashed key from the SHA-256 function, we use this hash to figure out the starting conditions for the Lorenz system \cite{lorenz1963deterministic}. These starting conditions will determine how the chaotic sequence used for encryption acts.
	
	The initial conditions \( x_0, y_0, z_0, w_0 \) are computed using the combination of XOR operations and summation. This process is inspired by Li et. al. \cite{liarticle}. For each initial condition, we take specific chunk and apply the following calculations:
	
	\begin{equation}
		x_0 = f(K_{17} - K_{20})
	\end{equation}
	\begin{equation}
		y_0 = f(K_{21} - K_{24})
	\end{equation}
	\begin{equation}
		z_0 = f(K_{25} - K_{28})
	\end{equation}
	\begin{equation}
		w_0 = f(K_{29} - K_{32})
	\end{equation}
	
	For example, \( f \) function for \(x_0 \) is:
	\begin{equation}
		\begin{split}
			x_0 = & \left( (k_{17} \oplus k_{18}) \oplus (k_{18} \oplus k_{19}) \oplus (k_{19} \oplus k_{20}) \right) \\
			& \times \left( \sum_{j=1}^{32} \frac{k_j \times 2^{j-1}}{2^{47}} \right)
		\end{split}
	\end{equation}

	These operations ensure that even a tiny changes in the key will result in significantly different initial conditions. Thus, the encryption process is very sensitive to the secret key.

	 \begin{figure}[h!]
		     \centering
		     \includegraphics[width=0.25\textwidth]{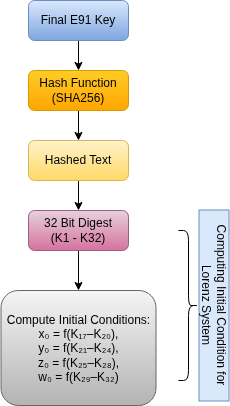} % Replace with the path to the image
		     \caption{Key Hashing and Initial Conditions Derivation Process}
		     \label{fig:key_hashing_initial_conditions}
		 \end{figure}

	\subsection{Generate Chaotic Sequence}
	
	The generation of chaotic sequences begins with initial conditions \( (x_0, y_0, z_0, w_0) \) derived from the SHA-256 hash of the encryption key. These initial conditions are crucial, as small changes lead to drastically different chaotic sequences. The Lorenz system is solved using the RK45 solver (Runge-Kutta method with adaptive step size) from time \(t = 0\) to \(t = T + M \times N\), where \(T\) is the transient time, and \(M \times N\) represents the image dimensions.
	
	The first \(T\) steps are discarded to allow the system to stabilize, after which valid chaotic sequences \(x(t), y(t), z(t), w(t)\) are obtained. These continuous sequences are then normalized and scaled to fit within the pixel range of 0 to 255, suitable for encryption purposes.
	 \begin{figure}[h!]
		     \centering
		     \includegraphics[width=0.7\textwidth]{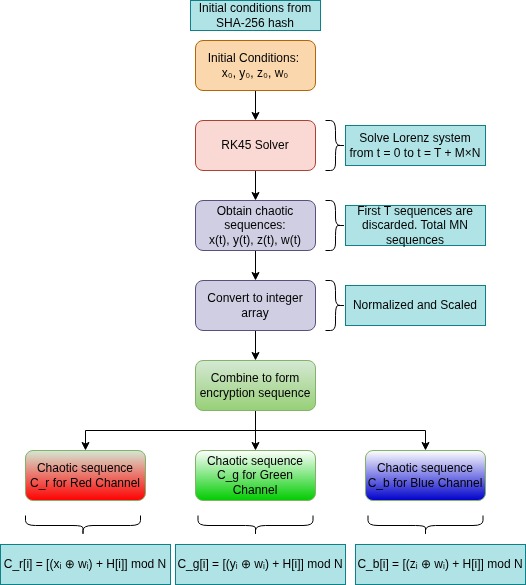} % Replace with the path to the image
		     \caption{Chaotic Sequence Generation Process}
		     \label{fig:Chaotic Sequence Generation}
		 \end{figure}
	Then the chaotic sequences are used to generate encryption sequences for the red, green, and blue channels as follows:
	\begin{equation}
		C_r = \left[ (x \oplus w) + H \right] \mod N
	\end{equation}
	\begin{equation}
		C_g = \left[ (y \oplus w) + H \right] \mod N
	\end{equation}
	\begin{equation}
		C_b = \left[ (z \oplus w) + H \right] \mod N
	\end{equation}
	where \(C_r, C_g, C_b\) represent the encryption sequences for the red, green, and blue channels, respectively, \(x, y, z, w\) are the chaotic values and \(H\) is the value derived from the hashed key. These chaotic sequences are then used to encrypt the image.

	\subsection{Encryption}
	The image encryption process involves multiple steps, with key hashing, initial condition generation, and chaotic sequence generation being the key stages. Reservoir Computing and random patterns are used to protect the image by applying chaotic transformations to its color channels.
	
	\begin{figure}[h!]
		\centering
		\includegraphics[width=0.8\textwidth]{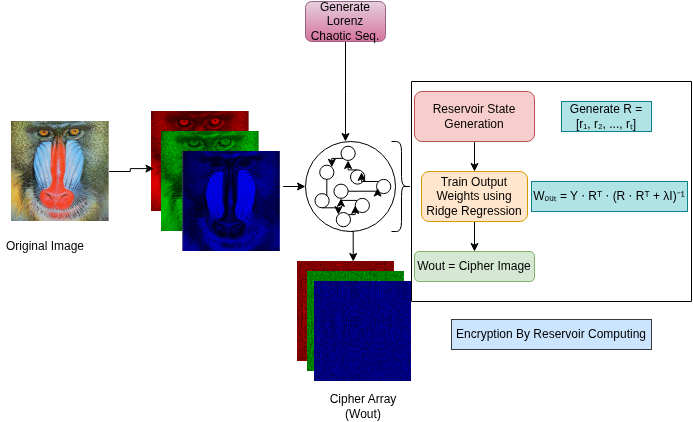} 
		% Include the path to the encryption process diagram
		\caption{Image Encryption Process using Chaotic Sequences and Reservoir Computing}
		\label{fig:encryption_process}
	\end{figure}
	
	The Reservoir Computing model uses these chaotic sequences to transform the original image into the encrypted cipher image, ensuring robust security for the encryption. This process is called the separation mode process \cite{krisharticle}. The equations used in this process are as follows:
	
	\textbf{Reservoir State Update:}
	\begin{equation}
		\mathbf{r}_{t+1} = (1 - \alpha)\mathbf{r}_t + \alpha \cdot \tanh(\mathbf{W}_{in} \cdot \mathbf{u}_t + \mathbf{A} \cdot \mathbf{r}_t)
	\end{equation}
	\textbf{Training the Output Layer (Ridge Regression):}
	\begin{equation}
		\mathbf{W}_{out} = \mathbf{Y} \mathbf{R}^T (\mathbf{R} \mathbf{R}^T + \lambda \mathbf{I})^{-1}
	\end{equation}
	\subsection{Decryption}
	The cipher array \(W_{out}\) holds the encrypted image, which has red (\(W_{out_r}\)), green (\(W_{out_g}\)), and blue (\(W_{out_b}\)) channels that are decoded one at a time. The Reservoir Computing model and its accompanying chaotic sequence  \(D_r\), \(D_g\), and \(D_b\) are used to process each channel in order to decrypt it and get back the original pixel values.
	
	The same secret key that was used to encrypt the image data is also used to decode it. The model regenerates the reservoir states and uses an inverse transformation to make sure that the pixel values for each color channel are restored correctly. The three channels are put back together to make the full-color image after they have been decoded.
	 \begin{figure}[h!]
		     \centering
		     \includegraphics[width=0.8\textwidth]{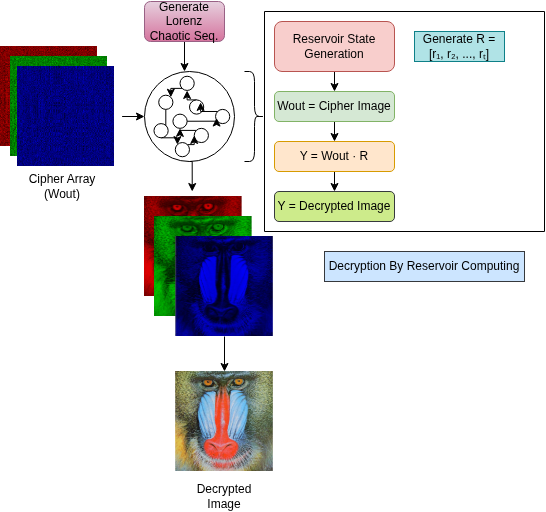} % Include the path to the decryption process diagram
		     \caption{Restoring the Original Image from the Encrypted Cipher Image}
		     \label{fig:decryption_process}
		 \end{figure}
	At last, the pixel values are adjusted to the 0–255 range to prevent overflow or underflow, producing a valid image. This encryption approach is safe because the original image can only be perfectly recreated with the right key.
	The equation used to decrypt the image is:
	\begin{equation}
		\mathbf{Y} = \mathbf{W}_{out} \cdot \mathbf{R}
	\end{equation}
	
	\section{Result and Discussion}
	Here we performed performance test of the hybrid model to see how it performs, security analysis to observe it is secure or not, and finally visual analysis to see how it looks like after encryption and decryption.
	
	\subsection{Performance Analysis}
	This section presents a performance analysis that evaluates the effectiveness of the system using different network sizes, encryption and decryption times, and key generation rates.
	
	\subsubsection{\textbf{MAE of Decrypted Image in Different Network Sizes}}
	We examined the Mean Absolute Error (MAE) between the original and decrypted photos for different network sizes and image resolutions (Lena and Baboon at 256×256 and 512×512). 
	\begin{figure}[h!]
		\centering
		\includegraphics[width=0.5\textwidth]{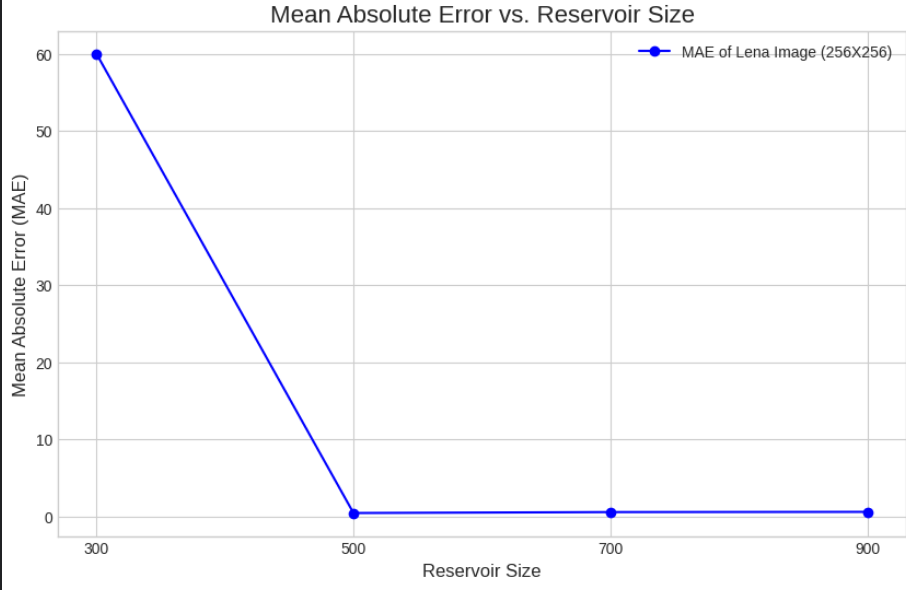} % Replace with the path to the image
		\caption{MAE vs Network size of Lena (256X256)}
		\label{fig:MAE vs Network Size (Lena 256)}
	\end{figure}
	\begin{figure}[h!]
		\centering
		\includegraphics[width=0.5\textwidth]{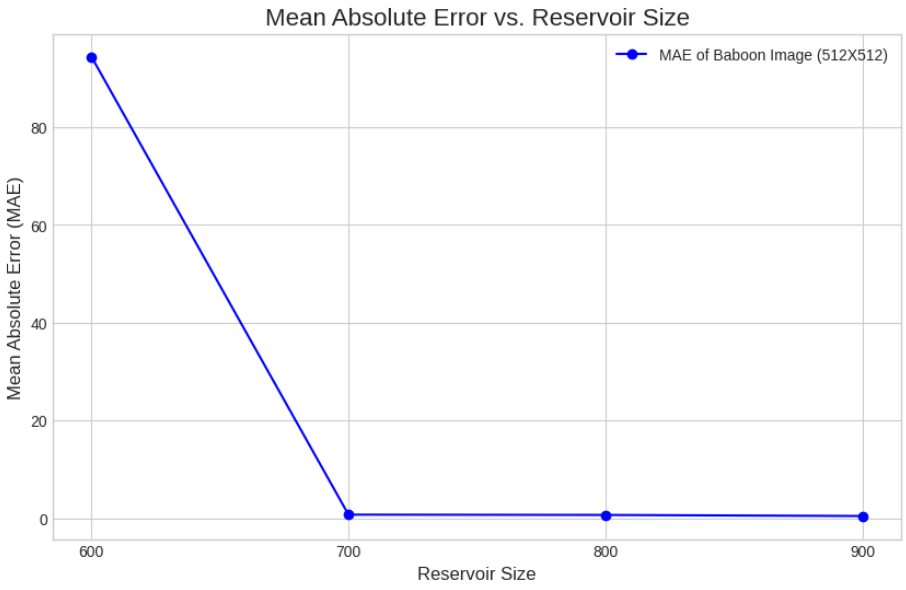} % Replace with the path to the image
		\caption{MAE vs Network size of Baboon (512X512)}
		\label{fig:MAE vs Network Size (Baboon 512)}
	\end{figure}
	The results suggest that larger networks reduce MAE significantly, making decryption better. The biggest changes happen when the network size is between 300 and 600. After that, the changes get smaller and smaller. Both pictures showed the same trends, which showed that larger networks make decryption more accurate.
	
	\subsubsection{\textbf{Encryption and Decryption Mean Time}}
	
	We examined how long it took to encrypt and decrypt images of different sizes and networks of different sizes. Table \ref{table: Mean time} displays the average time it took to encrypt and decrypt on test images. Each image was tested 100 times to get the average.
	
	\begin{table}[h!]
		\centering
		\small
		\resizebox{\textwidth}{!}{ % Resize the table to fit the text width
			\begin{tabular}{c c c c c}
				\rowcolor[HTML]{F4CCCC} 
				\textbf{Image Name} & \textbf{Image Size} & \textbf{Network Size} & \textbf{Mean Encryption Time(s)} & \textbf{Mean Decryption Time(s)} \\
				\hline
				Baboon & 512 & 600 & 0.3689 & 0.2799 \\
				Baboon & 256 & 500 & 0.1313 & 0.0843 \\
				Baboon & 128 & 300 & 0.0297 & 0.0164 \\
				Lena & 512 & 600 & 0.3515 & 0.2745 \\
				Lena & 256 & 500 & 0.1312 & 0.0878 \\
				Lena & 128 & 300 & 0.0296 & 0.0176 \\
			\end{tabular}
		}
		\caption{Mean Encryption and Decryption Times for Different Images, Image Sizes, and Network Sizes}
		\label{table: Mean time}
	\end{table}
	% \begin{table}[htbp]
		% \caption{Table Type Styles}
		% \begin{center}
			% \begin{tabular}{|c|c|c|c|}
				% \hline
				% \textbf{Table}&\multicolumn{3}{|c|}{\textbf{Table Column Head}} \\
				% \cline{2-4} 
				% \textbf{Head} & \textbf{\textit{Table column subhead}}& \textbf{\textit{Subhead}}& \textbf{\textit{Subhead}} \\
				% \hline
				% copy& More table copy$^{\mathrm{a}}$& &  \\
				% \hline
				% \multicolumn{4}{l}{$^{\mathrm{a}}$Sample of a Table footnote.}
				% \end{tabular}
			% \label{tab1}
			% \end{center}
		% \end{table}

	The results indicate that as image and network sizes decrease, both encryption and decryption times decrease as well. The encryption time for a 512x512 Baboon image with a 600 network size is 0.3689 seconds, which drops to 0.0297 seconds for a 128x128 image with a 300 network size. The network size directly impacts processing time, with larger networks requiring more time due to the increased number of neurons and weights. However, the system shows efficient performance even with larger images.
	
	\subsubsection{\textbf{E91 Protocol Key Generation Rate}}
	
	The E91 protocols' key generation rate, used in quantum cryptography, was tested with different singlet state numbers. The table \ref{tab: e91} shows the results for key length, key generation time, and key generation rate.
	
\begin{table}[h!]
	\centering
	\small
	\renewcommand{\arraystretch}{1.5} % Increase row height for better readability
	\begin{tabular}{c|c|c|c}
		\rowcolor[HTML]{F4CCCC} 
		\textbf{Singlet State Used} & \textbf{Key Length} & \textbf{Key Generation Time (s)} & \textbf{Key Generation Rate (bps)} \\
		\hline
		25  & 7    & 4.39  & 1.59  \\
		100 & 25   & 4.66  & 5.37  \\
		250 & 57   & 5.42  & 10.52 \\
		500 & 106  & 8.66  & 12.24 \\
		700 & 157  & 11.43 & 14.87 \\
		\hline
	\end{tabular}
	\caption{E91 Protocol Key Generation Rate}
	\label{tab: e91}
\end{table}
	
	It shows that increasing the number of singlet states leads to longer key lengths and faster key generation rates. Although the key generation time increases with more singlet states, the key generation rate improves, providing a stronger, faster protocol with more secure keys.
	
	\subsection{Security Analysis}
	
	In this part, we examine how well the proposed approach for encrypting images works. using key sensitivity, NPCR (Number of Pixel Change Rate), UACI (Unified Average Changing Intensity), and histogram analysis.
	\subsubsection{\textbf{CHSH Value Analysis}}
	The CHSH inequality confirms the E91 protocol’s secure quantum key distribution, with values in Table~\ref{tab:chsh} exceeding \( |S| > 2 \), ensuring entanglement-based protection against eavesdropping.
	\begin{table}[h!]
			\raggedright % Align content to the left
			\caption{CHSH Values for Image Encryption Configurations} % Caption will follow the minipage alignment
			\footnotesize % Smaller font size
			\resizebox{\textwidth}{!}{ % Resize within the minipage width
				\begin{tabular}{|c|c|c|c|c|}
					\hline
					\textbf{Image Name} & \textbf{Image Size} & \textbf{Singlet States} & \textbf{Key Length} & \textbf{CHSH Value (\( S \))}\\
					\hline
					Baboon & 512 & 500 & 106 & 2.82\\
					Baboon & 256 & 250 & 57 & 2.79\\
					Lena & 512 & 500 & 106 & 2.80\\
					Lena & 256 & 250 & 57 & 2.78\\
					\hline
				\end{tabular}
			}
			\label{tab:chsh}
	\end{table}
	\subsubsection{\textbf{Key Sensitivity}}
	
\begin{figure}[h!]
		\centering
		\includegraphics[width=0.8\textwidth]{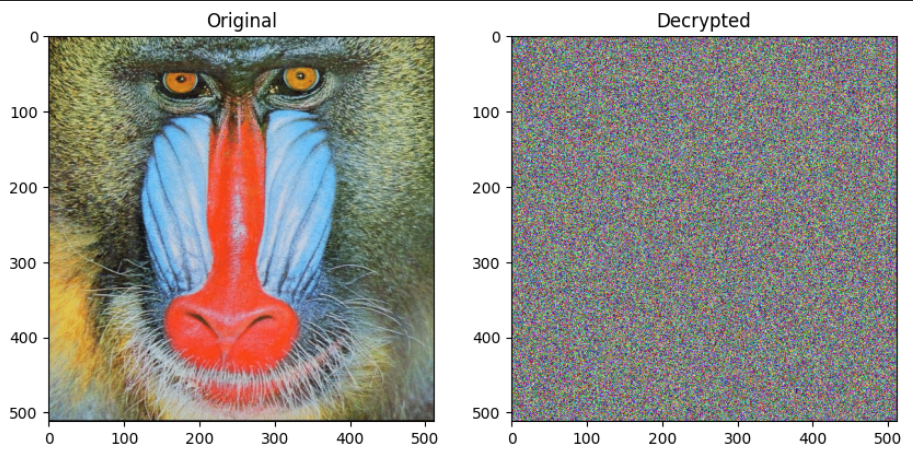} % Replace with the path to the image
		\caption{Decryption Using the Slightly Changed Key}
		\label{fig:Key Sensitivity}
	\end{figure}

	Its strong key sensitivity makes the encryption mechanism more secure. Fig \ref{fig:Key Sensitivity} shows that a change of only one bit in the key creates a very different hash and decrypted image. The big difference between the original image and the changed image shows how well the system can stop illegal access, which proves that it is reliable.
	
	\subsubsection{\textbf{NPCR and UACI}}
	
	We employ both NPCR and UACI to check how strong and high-quality the encrypted images are when the keys change. The NPCR results reveal that even little changes to the source image cause big changes in the encrypted image, with NPCR values always being above 99.6\%. Also, UACI values show that the hybrid encryption process does a good job of mixing up the pixel values, which makes the encrypted image look random and very hard to decrypt.
	
	\begin{table}[h!]
		\caption{NPCR and UACI for Different Images and Sizes}
		\centering
		\small
		\renewcommand{\arraystretch}{1.5} % Increase row height for better readability
		\begin{tabular}{|c|c|c|c|}
			\hline
			\textbf{Image Name} & \textbf{Image Size} & \textbf{NPCR} & \textbf{UACI}\\
			\hline
			Baboon & 512 & 99.64\% & 0.4989\\
			Baboon & 256 & 99.76\% & 0.4936\\
			Baboon & 128 & 99.82\% & 0.4958\\
			Lena & 512 & 99.64\% & 0.4900\\
			Lena & 256 & 99.78\% & 0.4968\\
			Lena & 128 & 99.80\% & 0.4976\\
			\hline
		\end{tabular}
	\end{table}
	
	% \begin{table}[h!]
		% \caption{UACI for Different Images and Sizes}
		% \centering
		% \small
		% \renewcommand{\arraystretch}{1.5} % Increase row height for better readability
		% \begin{tabular}{|c|c|c|}
			% \hline
			% \textbf{Image Name} & \textbf{Image Size} & \textbf{UACI} \\
			% \hline
			% Baboon & 512 & 0.4989 \\
			% Baboon & 256 & 0.4936 \\
			% Baboon & 128 & 0.4958 \\
			% Lena & 512 & 0.4900 \\
			% Lena & 256 & 0.4968 \\
			% Lena & 128 & 0.4976 \\
			% \hline
			% \end{tabular}
		% \label{tab: UACI}
		% \end{table}

	\subsubsection{\textbf{Histogram Analysis}}
	
\begin{figure}[ht]
		\centering
		\includegraphics[width=0.8\textwidth]{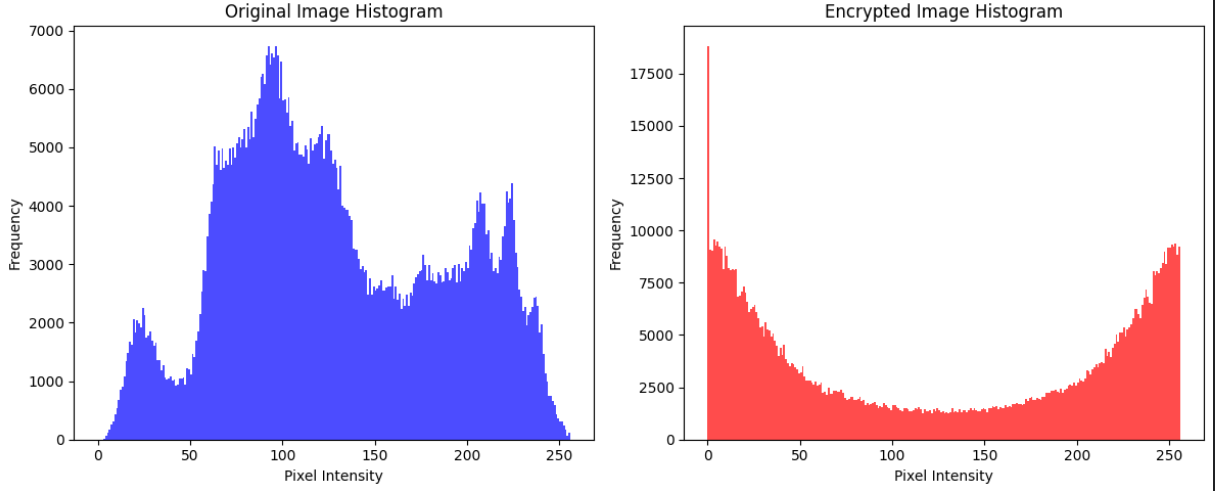}
		\caption{Histogram comparison of the original and encrypted images}
		\label{fig: hist}
	\end{figure}

	We utilize histogram analysis to look at how the pixel intensity is spread out in encrypted photos. The histogram of the original image usually reveals a pattern in the intensities of the pixels, while the histogram of the encrypted image shows a more even distribution. This means that the encryption process has done a good job of randomizing the pixel values. This even distribution is good because it makes the encrypted image look like random data, which protects it from frequency analysis and statistical attacks.
	Fig \ref{fig: hist} illustrates the histogram of the original and encrypted image, which are very different from each other. The histogram of the original image has definite peaks that show the pixel intensities, whereas the histogram of the encrypted image is equally spaced out. This shows that the encryption procedure worked to keep the image safe.
	
	Overall, the results from NPCR, UACI and histogram analysis describes that the proposed method ensures strong security by effectively randomizing the encrypted image and making it resistant to various kind of attacks.
	
	\subsection{Visual Analysis}
	By observing the input image, encrypted image, and decrypted image next to each other, we can see how perfect the encryption method works. We can find out how greatly the hybrid system help to protect visual information. The fig \ref{fig:lenavisu} below show the original image (left one) then the encrypted image (middle one) and lastly decrypted image next to each other.
	
	\begin{figure}[ht]
		\centering
		\includegraphics[width=0.8\textwidth]{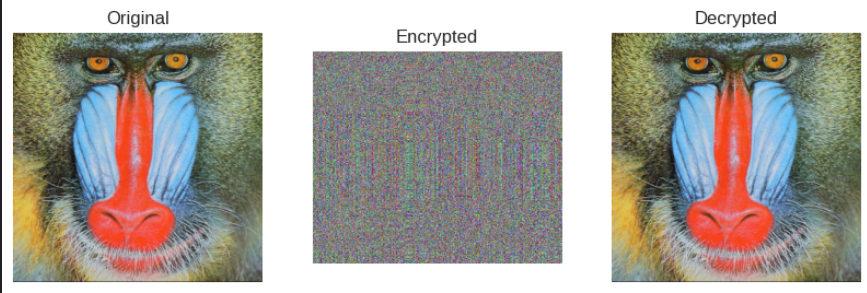} 
		\caption{Orginal, Encrypted and Decrypted Image}
		\label{fig:lenavisu}
	\end{figure}
	
	% \begin{figure}[ht]
		% \centering
		% \includegraphics[width=0.5\textwidth]{images/babVisu.png}
		% \caption{Orginal, Encrypted and Decrypted Image of Baboon (512X512)}
		% \label{fig:babVisu}
		% \end{figure}
	The visual tests show that the encryption approach works well.  The encrypted image seems completely random, so you can't see any of the original pixels.  The decryption brings back the original image fully, with no loss.
	\section{Discussion and Findings}
	The question this research tried to answer is whether quantum cryptography, chaos theory, and reservoir computing could all be used together to protect pictures. The study found that putting these different technologies together can make security systems which are safe and work well a lot better.
	It has been proven that the Quantum Key Distribution (QKD) protocol's E91 protocol can protect encryption keys from quantum dangers. It is used to make random sequences for image encryption. The Lorenz system is an example of a chaotic system. This makes it very hard to break with traditional cryptanalysis. Reservoir computing is a very significant part of making the process perform better.
	It is a key aspect in encrypting images with high resolution in real time. Quantum cryptography, chaos theory, and reservoir computing could all work together to make encryption systems that are exceedingly safe, fast, scalable and efficient.
	\section{Conclusion and Future Work}
	This study looked into a process for encrypting images that combines quantum key distribution or quantum cryptography, chaos theory and reservoir computing. The results showed that the hybrid technique is a safe, fast, and effective way to encrypt data. Still, more work needs to be done to smoothly fit these technologies into a useful framework. More study into system optimization, scalability, and quantum error correction could lead to big improvements in image encryption. We can create a strong, fast, and scalable encryption system that can handle both conventional and quantum threats by getting beyond these problems.
	
	This study looked at the theoretical and practical effects of combining quantum key distribution (QKD), chaos theory, and reservoir computing. However, more research should be done on scalability and real-time encryption to make sure that large images and videos can be handled quickly and efficiently, while also addressing computational needs and latency issues. Also, combining Quantum Error Correction (QEC) \cite{qecCampbell:2024ssp} with the E91 protocol could make it more resistant by reducing quantum noise and decoherence, which would make sure that encryption is still secure even if there are transmission mistakes. These improvements would make things more useful and safe.
	\newpage
	\bibliographystyle{ieeetr}
	\bibliography{ref}
\end{document}